# HIGH-BRIGHTNESS BEAMS FROM A LIGHT SOURCE INJECTOR: THE ADVANCED PHOTON SOURCE LOW-ENERGY UNDULATOR TEST LINE LINAC[*]

G. Travish, S. Biedron, M. Borland, M. Hahne, K. Harkay, J. W. Lewellen, A. Lumpkin,
S. Milton, N. Sereno, Advanced Photon Source, Argonne National Laboratory, Argonne, IL 60439

*Abstract*

The use of existing linacs, and in particular light source injectors, for free-electron laser (FEL) experiments is becoming more common due to the desire to test FELs at ever shorter wavelengths. The high-brightness, high-current beams required by high-gain FELs impose technical specifications that most existing linacs were not designed to meet. Moreover, the need for specialized diagnostics, especially shot-to-shot data acquisition, demands substantial modification and upgrade of conventional linacs. Improvements have been made to the Advanced Photon Source (APS) injector linac in order to produce and characterize high-brightness beams. Specifically, effort has been directed at generating beams suitable for use in the low-energy undulator test line (LEUTL) FEL in support of fourth-generation light source research. The enhancements to the linac technical and diagnostic capabilities that allowed for self-amplified spontaneous emission (SASE) operation of the FEL at 530 nm are described. Recent results, including details on technical systems improvements and electron beam measurement techniques, will be discussed. The linac is capable of accelerating beams to over 650 MeV. The nominal FEL beam parameters used are as follows: 217 MeV energy; 0.1–0.2% rms energy spread; 4–8 µm normalized rms emittance; 80–120 A peak current from a 0.2–0.7 nC charge at a 2–7 ps FWHM bunch.

## 1 OVERVIEW

The low-energy undulator test line (LEUTL) self-amplified spontaneous emission (SASE) free-electron laser (FEL) project has as its primary goal the identification and study of issues relevant to linac-based fourth-generation x-ray light sources [1], including verifying the behavior of the SASE FEL with varying electron beam parameters. Therefore, good characterization and control of the electron beam is critical to the success of the project.

The LEUTL project has taken a conservative approach towards producing a drive beam for the SASE FEL. The project began with the available APS injector and made numerous improvements and additions to the system over the past 2+ years to allow for the production, preservation, and measurement of high-brightness beams. A description of the system, starting at the head of the linac, follows.

In collaboration with the Accelerator Test Facility (ATF) at Brookhaven National Laboratory (BNL) a 1.6-cell, S-band photocathode (PC) rf gun and emittance compensation solenoid [2] was installed at the near-optimal drift distance to the existing linac. The photoinjector replaced an existing low peak current, large energy spread DC thermionic gun. The ATF PC gun has been well characterized and proven in several installations around the world, and is intended to produce beams of over 100-A peak current with emittances below 5 µm, as opposed to the DC gun that produced < 75-A peak current and > 100-µm emittances.

A drive laser for the photocathode gun was purchased commercially, installed into a laser room constructed adjacent to the linac tunnel, and integrated into the APS control system [3]. An optical transport line delivers the laser either into the APS linac enclosure or into an rf test area adjacent to the laser room.

The APS injector linac is constructed from standard SLAC-type traveling-wave disk-loaded linac sections. One linac section immediately following the gun is powered by a single klystron; the remaining twelve linac sections are grouped into three sectors, each of which is powered by a single klystron equipped with SLED cavities. Two thermionic-cathode rf guns provide redundant injection capability for the APS storage ring [4], and one of the guns has performance sufficient to allow limited FEL studies. Improvements made to the linac for LEUTL (as well as APS operations) include rearranging the focusing lattice and realigning all linac sections and magnets. Upgrades made to the rf system improved the power stability from over 7% (pk-pk over several seconds) to better than 2%, and phase stability from several degrees to about 2 degrees over short time periods (seconds). Further improvements to the focusing lattice and the rf system are ongoing.

The APS linac and LEUTL transport lines contain four energy spectrometers, a three-screen emittance measurement region, and several metal foils for providing light to optical transition-radiation (OTR) and coherent transition-radiation (CTR) diagnostics. The undulator hall proper contains additional diagnostics stations for both the electron beam and the photon beam generated by the FEL interaction. The linac diagnostics are described in detail in section 4, but first we discuss the project goals and design parameters.

## 2 PROJECT GOALS

The LEUTL FEL operates in high-gain SASE mode, generating light from 530 nm down to less than 100 nm.

A principal portion of the FEL studies is being devoted to experimental verification of scaling laws derived from SASE FEL theory, such as the variation of output power with beam parameter changes. Theoretical

---

[*] Work supported by the U.S. Department of Energy, Office of Basic Energy Sciences, under Contract No. W-31-109-ENG-38

comparison requires both high-resolution beam characterization and good beam stability over long and short time scales. Significant technical effort and study time are dedicated to improving and understanding the linac in order to achieve the above goals.

The technical development program also focuses on additional aspects of accelerator technology anticipated to be useful for future light source construction. As an example, a magnetic bunch compression system [5] has been installed into the linac with parameters similar to the BC-1 compressor in the LCLS design [6]. The LEUTL compressor will allow for the study of the onset of coherent synchrotron radiation and other effects as the compression is varied. The bunch compressor will soon be used along with the LEUTL SASE FEL to verify the theoretical tradeoffs between high peak currents and emittance.

## 3 DESIGN PARAMETERS

The necessary electron beam parameters are straightforward to calculate from the undulator specifications (a planar undulator with a period of 3.3 cm, a normalized field of 3, and an effective average betafunction of 1.5 m at 217 MeV) and the desired FEL characteristics (maximum gain at 530 nm). Following the conservative approach described in the previous section, the initial demands on the electron beam were kept to within state of the art (see Table 1).

Table 1: Design beam parameters for the LEUTL FEL

| Parameter | Value |
| --- | --- |
| Energy | 217 MeV |
| Energy Spread (rms) | 0.1% |
| Charge | 1 nC |
| Bunch Length (FWHM) | 5–7 ps |
| Peak Current | 150 A |
| Emittance (normalized rms) | 5 µm |

Since achieving the performance listed in Table 1 is nontrivial, a number of diagnostics were installed to properly characterize the beam and provide guidance for improvements.

## 4 DIAGNOSTICS

Prior to the LEUTL project, the APS linac was only required to provide a low-brightness, modest current beam for injection [7]. As such, the linac diagnostics consisted of modest resolution devices used for tuning up the beam. After significant upgrade efforts, the APS linac now incorporates many diagnostics for determining various beam parameters throughout the system. The diagnostics used for characterizing the LEUTL beam are described below in roughly the order they appear on the beamline:

An Integrating Current Transformer (ICT), located just after the PC gun, measures beam charge; resolution is better than 10 pC, and accuracy is estimated to be 10% at nominal charges (1 nC).

YAG screens with charge-injection device (CID) cameras are located after the PC gun and at a few points along the accelerator. The screens are used to determine beam size, with an approximate resolution and accuracy of 50 µm (at present charge densities).

Beam position monitors (BPMs) provide beam trajectory information along the linac and transport line with a resolution of 50 µm (25 µm with improved electronics, better than 10 µm with the BPMs used in the transport lines after the linac). The BPM data are used in a trajectory control law (that employs a previously measured response matrix of steering magnets and BPMs) to actively correct the beam's average trajectory. The control law is especially useful in maintaining a trajectory that best minimizes and compensates for transverse wakes.

Bending magnets located in linac sections L3 (midway) and L5 (at the end of the linac) are used for determining beam energy and energy spread at ~135–160 MeV and ~217–650 MeV, with resolutions of 0.033%/pixel and 0.05%/pixel, respectively. Additional bend spectrometers are installed midway in the long transport line between the linac and undulators, as well as after the undulators.

Bunch length measurements can be made in one of three ways. One method uses an insertable mirror at the end of the linac to generate OTR that is transported to a streak camera with a resolution of about 1 ps. Bunch length is also measured using the last accelerating section (L5) and the so-called zero phasing technique [8]. Finally, an optical high resolution spectrometer applied to the FEL output light has been used to indirectly obtain the bunch length [9].

Emittance measurements are primarily made using a three-screen arrangement, with drift spaces between the screens. The three-screen system is accurate, helps with betafunction matching, and does not suffer from space charge effects or quadrupole calibration errors. A software utility automates the entire process. Presently the system uses YAG screens located several meters after the linac, but OTR mirrors may be employed in the near future to avoid saturation effects and improve resolution. While it is difficult to estimate the accuracy of the system, a resolution of better than 1 µm in emittance is possible when the beam is properly matched.

The above diagnostics were the primary tools used to characterize the photoinjector-generated beam. In addition, fluorescent screens and CID cameras are placed along the linac/LEUTL line. These are used in conjunction with a digitizing image analysis system to measure the beam's position and transverse profile. Finally, diagnostics on the drive laser measure pulse energy (joulemeter with a resolution of better than 1 µJ out of 500 µJ), spot size (CCD camera with a resolution and accuracy of 100 µm with improvements planned), and pulse length (single-shot autocorrelator with a resolution of better than 100 fs and a total system accuracy of perhaps 20%) on a shot-to-shot basis.

## 5 MEASURED PERFORMANCE

Using the above described diagnostics, the beam is well characterized each run shift, often several times per shift. The measurements are made concurrently when possible (i.e., charge and emittance), and in rapid succession when necessary (i.e., energy spread then emittance). Moreover, the measurements tabulated here represent repeatable data taken under the same linac conditions used to run the FEL. While the capability of saving and restoring machine settings has long existed at the APS [10], the performance levels demanded of the linac by LEUTL required substantial improvements in readback systems, power supplies, magnet degaussing routines, and rf system controls. Perhaps the most significant improvement made to the system over the past two years is this ability to restore and reproduce beam characteristics with minimal manual adjustments.

Referring to Table 2, the energy spread measurements are perhaps the least certain due to optical resolution limits in our present spectrometers. The bunch length measurements are typically made using the zero-phasing technique mentioned above (and agree with the less frequently used streak measurements). Finally, the emittance measurements have the largest variation due to trajectory jitter and wakefield effects, as we discuss in the next section.

Table 2: Measured parameters for the LEUTL beam when tuned for FEL operation

| Parameter | Value | Diagnostic |
|---|---|---|
| Energy | 217 MeV | Bend magnets |
| Energy Spread (FWHM) | < 0.15% | Bend magnets |
| Charge | 0.5 nC | ICT |
| Bunch Length (FWHM) | 3–4 ps | Zero phase |
| Peak Current | ~140 A | *Calculated* |
| Emittance (norm. rms) | < 7 μm | 3 screen |

## 6 ISSUES

Having followed a conservative and proven approach with the photoinjector, it was originally thought that the moderately high-brightness beams required for LEUTL would be straightforward to produce. However, due to the inherent sensitivity of the photoinjector to the solenoid field, laser spot size, and input rf, it became clear that simply duplicating past efforts was insufficient. Indeed, a number of constraints specific to LEUTL implied that a suitable operating regime would have to be discovered. Finding a good operating point for the photoinjector has been complicated by the need to traverse a number of accelerator structures before reaching a complete set of diagnostics. Below are some of the major issues that limited beam quality during the operating periods presented in this paper.

**Cathode nonuniformity:** While scanning the laser across the cathode surface, it was observed that some structure, as projected by the beam onto a YAG screen, remained stationary. These observations imply that the structure arose from some nonuniformity of the cathode, not the laser. Visual inspection using a 70 degree (off normal) cathode port supports the above observations: the cathode center appears pitted and concentric machining rings are evident.

**Wakefields:** Proper compensation of the transverse wakefield is critical to preserving the beam emittance through the long linac. Compensation involves finding the correct trajectory through the linac. Typically, wakefield compensation is used; however, the jitter and drifts in the various systems means that the transverse wakefields are only partially compensated and for only a portion of the shots. In addition to the above challenges, it is felt that the first linac section (right after the gun) has an undetermined defect that makes it impossible to find a trajectory that does not suffer from severe phase steering. Longitudinal wakefields are compensated by simple rf phasing, and have proven to be much less problematic than the transverse wakes.

**Jitter:** Transverse trajectory jitter (pointing error) greater than the beam diameter is often observed at the end of the linac. Most of the jitter appears to come from the injector. A significant portion of the jitter may be caused by the pointing jitter of the drive laser (which at the time of the measurements reported here was not relay imaged onto the cathode). The remainder of the jitter is caused by rf amplitude and phase jitter in combination with the possible linac section problem described above. Appropriate rf power and phase levels can be varied to find a "sweet spot" where the effects of jitter are reduced; however, the appropriate parameter set changes with beam charge, launch phase, emittance, etc.

## 7 NEXT STEPS

Driven by the dictates of the FEL and by work of interest to future fourth-generation light-sources, the LEUTL photoinjector will be pushed to produce lower-emittance beams with higher peak currents. The magnetic bunch compressor—currently undergoing commissioning—will further increase the peak current.

By reducing the jitter and drift of the various support systems (rf, power supplies, laser, etc.), a more optimal operating regime should be found. An improved resolution energy spectrometer is also being designed. Longer term improvements in the rf and laser subsystems are also being considered.

While the initial lasing goal of the LEUTL FEL has been met [11], the primary goal is to understand the FEL SASE process. As such, characterizing the input electron beam will continue to be paramount.

## REFERENCES

[1] S.V. Milton, Proc. EPAC2000, to be published.
[2] M. Babzien et al., Phy. Rev. E **57**, 6093 (1998).
[3] G. Travish et al., Proc. FEL 1999, to be published.
[4] J.W. Lewellen, Proc. PAC 1999, 1979, (1999).
[5] M. Borland et al., "A Highly Flexible Bunch Compressor for the APS LEUTL FEL," these proceedings.
[6] LCLS Design Study Report, SLAC-R-521, 1998.
[7] M. White et al., Proc. PAC 1995, 1073 (1996).
[8] D.X. Wang, Proc. LINAC 1996, 303, (1997).
[9] V. Sajaev et al., Proc EPAC2000, to be published.
[10] R. Soliday et al., "Automated Operation of the APS Linac Using the Procedure Execution Manager," these proceedings.
[11] S.V. Milton et al., Phy. Rev. Lett. **85**, 988 (2000).